\documentclass[prl,amssymb,twocolumn]{revtex4}

\usepackage{graphicx}
\usepackage{amsmath}

\begin{document}

 \newenvironment{myquote}[1]%
  {\list{}{\leftmargin=#1\rightmargin=#1}\item[]}%
  {\endlist}

\title{Quantum Relativity}
\author{Michael Spanner}
\affiliation{National Research Council of Canada, 100 Sussex Drive, Ottawa, Canada}

\begin{abstract}
Starting with a consideration of the implication of Bell inequalities in
quantum mechanics, a new quantum postulate is suggested in order to
restore classical locality and causality to quantum physics: only the
relative coordinates between detected quantum events are valid
observables.  This postulate supports the EPR view that quantum
mechanics is incomplete, while also staying compatible to the Bohr view
that nothing exists beyond the quantum.  The new postulate follows from
a more general principle of quantum relativity, which states that only
correlations between experimental detections of quantum events have a
real classical existence.  Quantum relativity provides a
framework to differentiate the quantum and classical world.
\end{abstract}

\date{\today}

\maketitle

\section{Bell Inequalities in Quantum Mechanics}

\begin{myquote}{0.15in}
``For in fact what is man in nature? A Nothing in comparison with the Infinite,
an All in comparison with the Nothing, a mean between nothing and everything."
\begin{flushright}
- Blaise Pascal
\end{flushright}
\end{myquote}

\begin{myquote}{0.15in}
``It's a combination of both.
I mean here is the natural instinct and here is control. You are to 
combine the two in harmony. [...] If you have one to the extreme you'll 
be very unscientific, if you have another to the extreme, you become
all of a sudden a mechanical man. No longer a human being. [...]
It is a successful combination of both. [...] So therefore it is not
pure naturalness or un-naturalness.  The ideal is:

\begin{center}
    un-natural naturalness or natural un-naturalness."
\end{center}

\begin{flushright}
- Bruce Lee
\end{flushright}
\end{myquote}

\begin{myquote}{0.15in}
\begin{center}
	``${\hat {x}}{\hat {p}} - {\hat {p}}{\hat {x}}=i\hbar$"
\end{center}
\begin{flushright}
- Heisenberg
\end{flushright}
\end{myquote}

\begin{myquote}{0.15in}
\begin{center}
	``$|x\rangle|p\rangle - |p\rangle|x\rangle$"
\end{center}
\begin{flushright}
- EPR
\end{flushright}
\end{myquote}

Einstein, Podolsky, and Rosen \cite{EPR} argued that quantum mechanics was
an incomplete description of physical reality.  Bohr \cite{Bohr} maintained that
there was nothing more beyond the quantum.  Bell \cite{Bell} proposed a
scenario that could test these perspectives.

\begin{figure}[]
	\centering
	\includegraphics[width=\columnwidth]{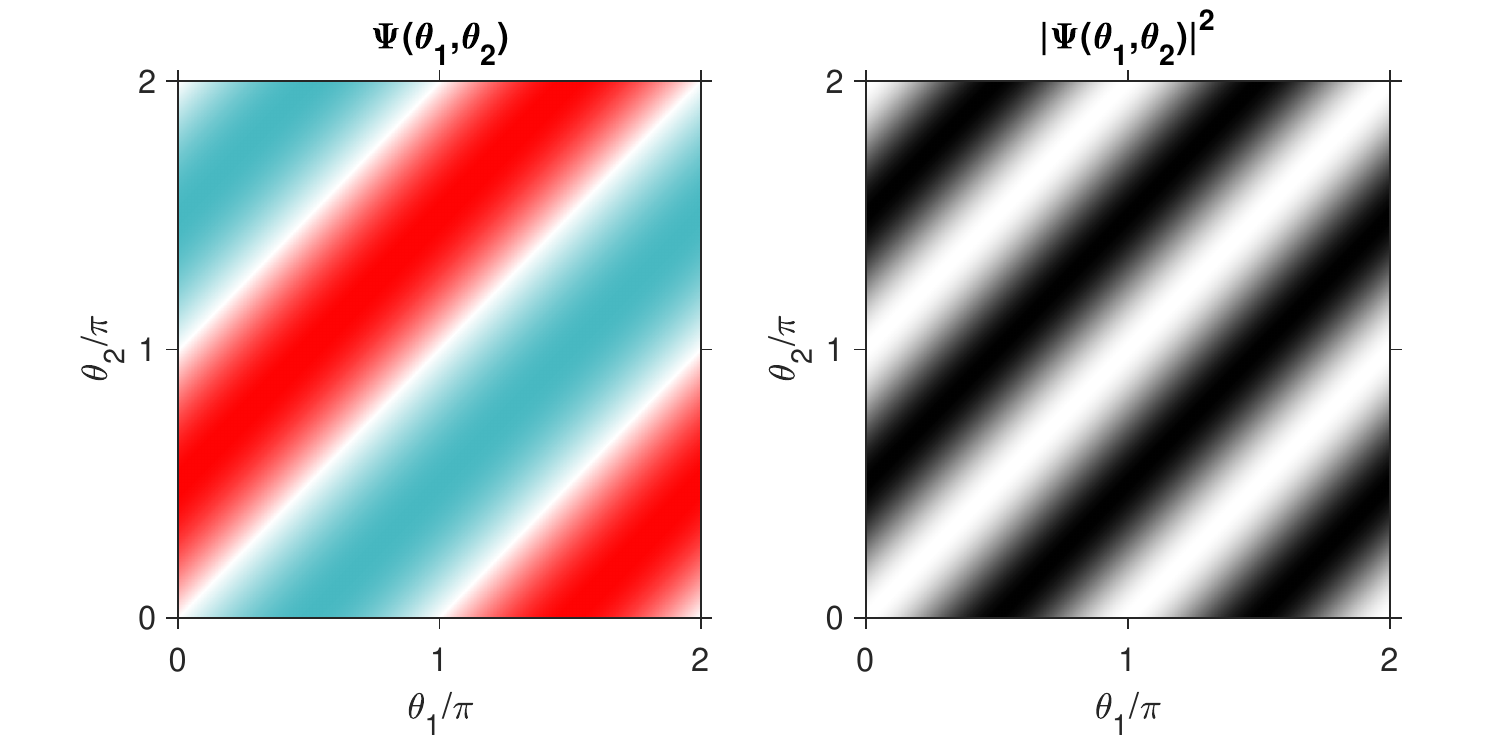}
	\caption{Left: Wavefunction of the singlet state from Eq.(\ref{EqBellState}).
	Blue is negative, red is positive.  Right: Correlated probability
	distribution of the singlet state from Eq.(\ref{EqBellDist}).
	White is zero, black is maximum.}
	\label{FigBell}
\end{figure}

Consider a traditional Bell scenario where some initial object with zero
angular momentum breaks apart into two fragments moving in opposite directions
along the same line.  Each fragment is a 2D rotor that can spin in the plane
perpendicular to the spatial motion.  The singlet state for this system usable
in a Bell scenario can be written as
\begin{equation}
	|\psi\rangle = |V\rangle|H\rangle - |H\rangle|V\rangle
\end{equation}
where the $|V\rangle$ and $|H\rangle$ states for a rotor are given by
\begin{equation}
	\langle \theta |V\rangle = \cos(\theta) , \:\:
	\langle \theta |H\rangle = \sin(\theta).
\end{equation}
Standard normalization of the wavefunctions is ignored, it does not matter for
the following discussion. These states are linear combinations of the $m=\pm 1$
angular momentum states, meaning that they are compatible with the idea that
which fragment received $\pm 1$ angular momentum is not known.  The double-angle 
wavefunction is given by 
\begin{eqnarray}
	\psi(\theta_1,\theta_2) &=& \langle\theta_1,\theta_2|\psi\rangle \nonumber \\
	                        &=& \cos(\theta_1)\sin(\theta_2) - \sin(\theta_1)\cos(\theta_2).
\end{eqnarray}
Now include the coordinate $\Delta\theta$ defined by 
$\theta_2=\theta_1+\Delta\theta$.  
With this coordinate the wavefunction becomes
\begin{eqnarray} \label{EqBellState}
	\psi(\theta_1,\Delta\theta) &=& \cos(\theta_1)\sin(\theta_1+\Delta\theta) \nonumber \\
	                            & &  - \sin(\theta_1)\cos(\theta_1 + \Delta\theta) \nonumber \\
	                            &=&  \sin(\Delta\theta).
\end{eqnarray}
The correlated probability distribution of detecting the fragments is then given by
\begin{equation}\label{EqBellDist}
	|\Psi(\theta_1,\Delta\theta)|^2 = |\Psi(\Delta\theta)|^2 = \sin^2(\Delta\theta).
\end{equation}
The wavefunction and distribution are plotted in Fig.\ref{FigBell}.  

As can be seen from the previous two equations or from the plots, the
wavefunction and correlated probability distribution depend only on the
relative angle $\Delta\theta$ between the coordinates at the isolated
detectors.  This is the core of the Bell inequalities.  The original Bell
inequality \cite{Bell} and the CHSH \cite{CHSH} version are procedures to
test for this property. I'll call the underlying symmetry, that the probability
of correlated detection only depends on the relative angle $\Delta\theta$ of
the two detections, the Bell correlation.  Assuming that the singlet state
represents two independent classical objects with individual hidden variables,
the specific symmetry of the corresponding classical probability distribution
can not reproduce the Bell correlation \cite{Bell}.

\section{Quantum Relativity and Elements of Classical Reality}

\begin{myquote}{0.15in}
``We do not describe the world we see, we see the world we can describe."
\begin{flushright}
	- Ren\'e Descartes
\end{flushright}
\end{myquote}

\begin{myquote}{0.15in}
``As for my own opinion, I have said more than once, that I hold space 
to be something merely relative, as time is, that I hold it to be 
an order of coexistences, as time is an order of successions."
\begin{flushright}
- Gottfried Leibniz
\end{flushright}
\end{myquote}

If the perspective is taken that only the single relative coordinate between
the two detection events {\it exists} for quantum events, then constructing a
classical distribution that exhibits the Bell correlation is
trivial---it would satisfy the Bell correlation by construction.  In order to
make sense of the Bell situation, one way forward then is to add to quantum
mechanics the additional postulate that only the relative coordinates between
experimental settings are valid observables when comparing correlated
measurements of quantum events. In this way, violations of Bell inequalities
arise because the quantum world has fundamentally less degrees-of-freedom for
the hidden variables than Bell had assumed.  He was averaging over additional
non-existent classical configurations in deriving the classical side of
the inequality.  This is the meaning of the Copenhagen interpretation---if you
can not measure it, it effectively does not exist.  In this sense, EPR
\cite{EPR} and Bohr \cite{Bohr} are both correct. There really is nothing
beyond the quantum, but quantum mechanics could be considered incomplete in
that it was missing a postulate.

The postulate can also be thought of as a quantum relativity principle where
only the relative coordinates of experimentally-detected quantum events have a
real classical existence.  This statement will be called ``weak" quantum
relativity.  A ``full" version will be introduced below.  Consider doing a Bell
experiment in a completely empty universe where nothing exists except the
detectors and the singlet state. Only the relative angle between the detection
events {\it could} exist in this universe since there is nothing else in it to
define another angle against.  Even if the two detectors are classically
connected to the entire universe, there still isn't a reference to measure
anything except a single relative coordinate.  Alternatively, think of
measuring the position and momentum of a quantum particle.  Detecting a 
quantum event at a single point on a position-sensitive screen 
gives you information about position, but absolutely no information about
momentum.  To be a full element of classical reality requires that a particle
have both position and momentum simultaneously, but a single detection point
clearly does not have a momentum.  Therefore, a single quantum event does not
carry a full element of classical reality.  

What then is a quantum event?  A single measured quantum event must be
correlated with a second quantum event in order to be a valid element of
classical reality.  What have been called quantum particles are not full
elements of classical reality. Rather, it is actually the correlations between
the quantum particles that must be considered as the full elements of reality
within the classical description of the world.  One pair of correlated quantum
events carries one element of correlation, and one element of correlation is
only one element of classical reality.

\section{Uncertainty Principles}

For clarity, it is useful to start with sound as an analogy.  Sound can be
represented as a single value that depends on time, $S(t)$, or as a single
value that depends on frequency, $\tilde S(f)$, but not both. Gabor showed
\cite{Gabor} how this property leads to an uncertainty principle between $t$
and $f$.  He believed that his treatment of acoustics was simply a curious
mathematical analogy and did not suggest a serious connection with quanta of
the atomic world.

Frequencies and times do not exist independent of each other.  We write music
as a series of pitches that occur at different times, but this is
not really what is happening.  Music is defined not by absolute pitch occurring
at different times, but more by relative pitches changes that occur at the
correct time intervals.  This is why any particular song is recognizable as
long as all the frequency intervals remain the same.  Transposing to a
different pitch does not change the song (though it might change the mood of
the song as you perceive it).  Likewise, any particular song will be recognizable if it is shifted
in time (though you might not be in the mood to perceive it at all times).

The analogous situation holds true for position and momentum of quantum point
particles.  The path of a point object is fully characterized by a single
coordinate, say $x(t)$ or $p(t)$, but not both. One can be derived from the
other as long as we are only interested in intervals between positions and/or
momenta at different points in time.  In analogy with the uncertainty
principle for sound, this leads to the existence of the Heisenberg
uncertainty principle between $x$ and $p$ of a quantum coordinate.  Uncertainty
relations arise when we attempt a description of reality that is overcomplete.
Only relative distances and relative momenta carry meaning when we measure the
world.

The Heisenberg uncertainty principle also follows from the recognition that we
are only really able measure positions and times of events, that we live in
space-time.  Consider measuring the momentum of an object.  To do so in
practice requires measuring some property of that object at two points in
space-time, say for example the center-of-mass coordinate at two different
times
\begin{equation}
	p(t) = M \frac{x_{cm}(t+\Delta t) - x_{cm}(t)}{\Delta t}
\end{equation}
where $M$ is the mass of the object.  When momentum is measured, it is really
relative distances and relative times that are being measured.  Position
and momentum are not independent of each other.

\section{Building the Classical world}

\begin{myquote}{0.15in}
``If you with to make an apple pie from scratch, 
you must first invent the universe." 
\begin{flushright}
- Carl Sagan
\end{flushright}
\end{myquote}

\begin{figure*}[]
	\centering
	\includegraphics[width=\textwidth]{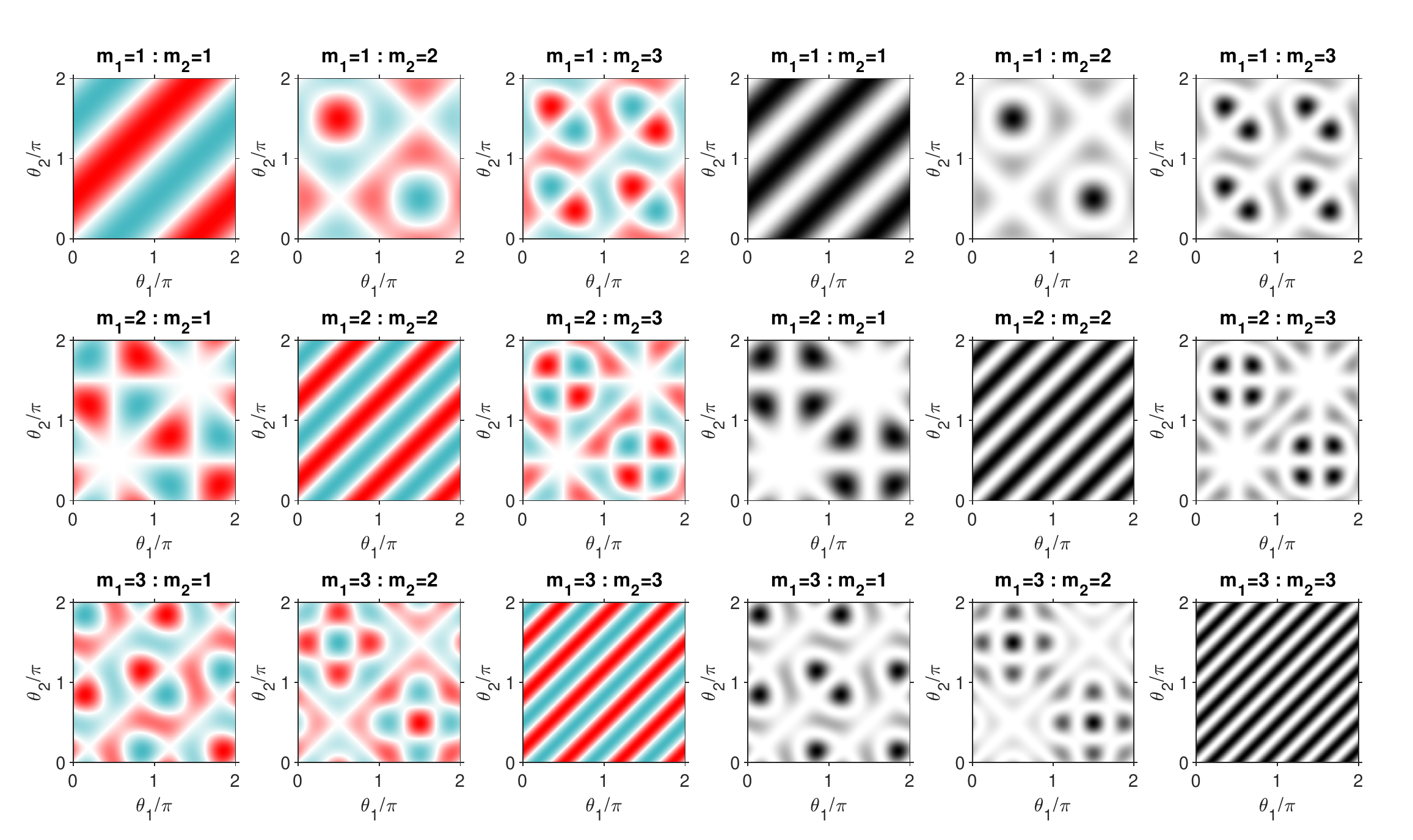}
	\caption{Wavefunctions (blue negative, red positive) and correlated
	probability distributions (white zero, black maximum) for a variety of
	negative excited-state singlets.}
	\label{FigNegative}
\end{figure*}

Can quantum relativity be reconciled with the apparent observation that
classical correlations, and not Bell correlations, seem to exist all around us?
That is, can the quantum relativity principle help us explain the quantum to
classical transition?  Let's invent the classical universe. 

Returning to the double-rotor system, allow the fragments to carry more than
one unit of angular momentum.  Let the system break apart from some unknown
initial state into the state 
\begin{eqnarray}\label{EqExcitedSinglets}
	\langle\theta_1,\theta_2|\psi\rangle &=& \cos(m_2\theta_1)\sin(m_1\theta_2) \nonumber \\
	                                     & & - \sin(m_1\theta_1)\cos(m_2\theta_2) 
\end{eqnarray}
where $m_{1}$ and $m_2$ are the total angular momenta of each fragment.  These
are like excited-state singlets.  A physical scenario that creates
excited-state singlets using spatial coordinates is presented below when
considering EPR-type dissociated diatomic scenarios.  The correlated
probability distribution for measurements of the two observed angles $\theta_1$
and $\theta_2$ is
\begin{eqnarray}\label{EqDistClassicalTransition}
	|\langle\theta_1,\theta_2|\psi\rangle|^2 &=& \big|\cos(m_1\theta_1)\sin(m_2\theta_2) \nonumber \\
	                                         & & -\sin(m_2\theta_1)\cos(m_1\theta_2)\big|^2.
\end{eqnarray}
The correlated distribution can not in general be written as a function of just
the relative angle $\Delta \theta$ for all combination of $m_1$ and $m_2$.
Consequently, some combinations of $m_1$ and $m_2$ will not show Bell
correlations nor be able to violate a Bell inequality.

To elaborate on this point and the symmetries that appear in the distributions
defined by Eq.(\ref{EqDistClassicalTransition}), Fig.(\ref{FigNegative}) shows
the double-angle wavefunctions and correlated distributions for a variety of
fragment momenta $m_1$ and $m_2$.  Two clear symmetries can be seen.  One set
has wavefunctions that can be written as a function of the single relative
coordinate $\Delta\theta$ with corresponding distributions that display Bell
correlations.  This set can violated of a Bell inequality.  The other set has
wavefunctions that can not be written as a function of a single relative
coordinate and distributions that do not display Bell correlations.  This
set can not violate a Bell inequality.  

From these two classes of symmetries, two classes of coordinates can be
recognized: quantum and classical.  Quantum coordinates arise when $m_1=m_2$
and display Bell correlations.  Classical coordinates arise when $m_1\neq m_2$
and do not have the Bell correlations.  The quantum or classical nature of the
coordinates being measured depends of the type of initial state prepared, 
and therefore the design of the experiment controls whether quantum
of classical effects can be seen in the measured coordinates.  

Note also that the symmetry of the wavefunctions is fermionic-like under
exchange of experimental coordinates.

\section{Positive Singlet States}

\begin{figure*}[]
	\centering
	\includegraphics[width=\textwidth]{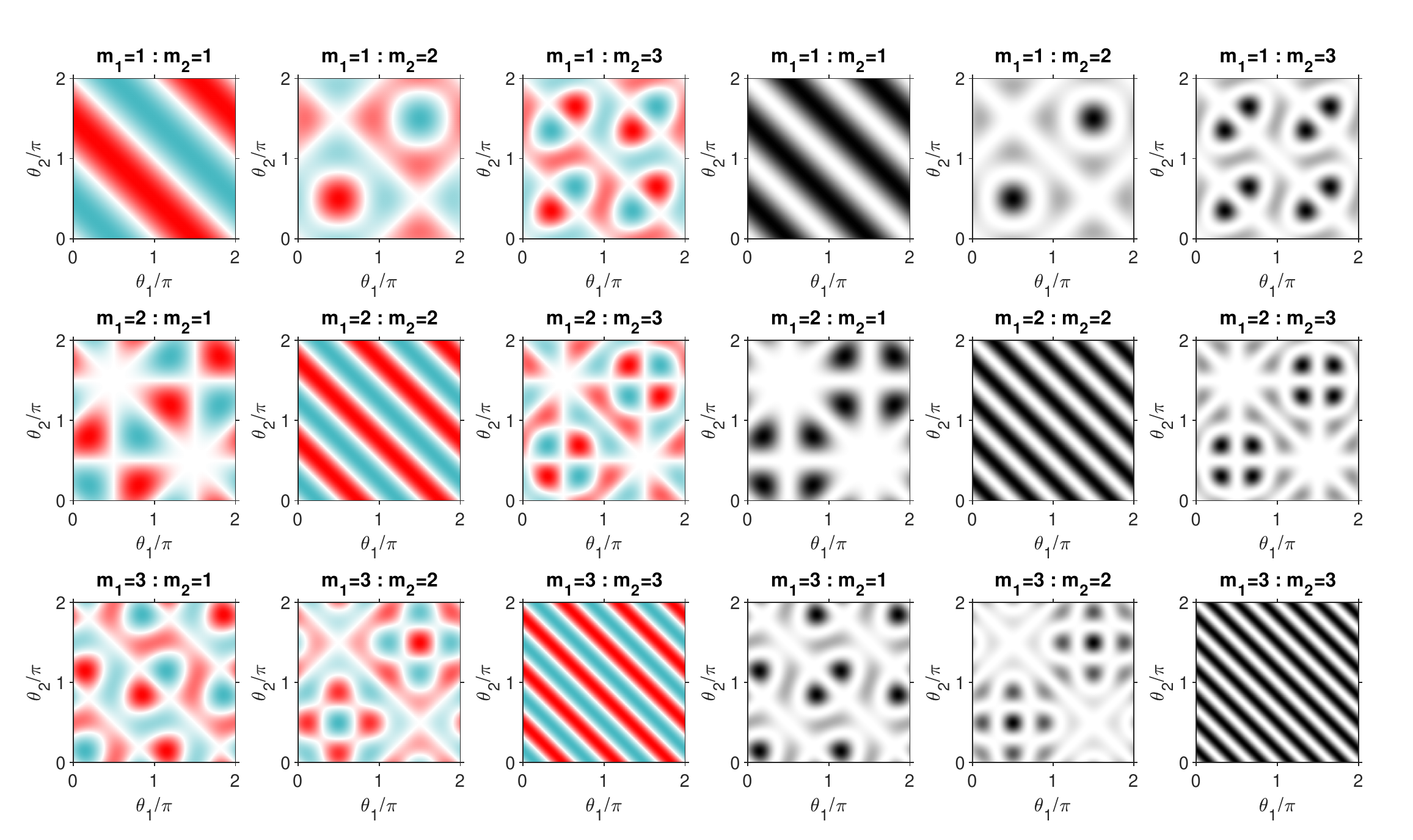}
	\caption{Wavefunctions (blue negative, red positive) and correlated
	probability distributions (white zero, black maximum) for a variety of
	positive excited-state singlets.}
	\label{FigPositive}
\end{figure*}

Instead of negative singlets $|A\rangle|B\rangle-|B\rangle|A\rangle$, 
consider now positive singlets $|A\rangle|B\rangle+|B\rangle|A\rangle$.
For the rotor system, the wavefunctions and correlated 
distributions of these states are
\begin{eqnarray}
	\langle\theta_1,\theta_2|\psi\rangle &=& \cos(m_1\theta_1)\sin(m_2\theta_2) \nonumber \\ 
	                                     & & + \sin(m_2\theta_1)\cos(m_1\theta_2)
\end{eqnarray}
and
\begin{eqnarray}
	|\langle\theta_1,\theta_2|\psi\rangle|^2 &=& \big|\cos(m_1\theta_1)\sin(m_2\theta_2) \nonumber \\ 
	                                         & & + \sin(m_2\theta_1)\cos(m_1\theta_2)\big|^2.
\end{eqnarray}
They are plotted in Fig.\ref{FigPositive}.  There are again classical-type
distributions appearing for $m_1\neq m_2$ that can not be factorized into a
single relative coordinate, but now the quantum-like $m_1=m_2$ states depend
only on the sum of the two experimental coordinates $\theta_{sum} =
\theta_1+\theta_2$ instead of $\Delta\theta$. I'll call this symmetry the
anti-Bell correlation.  This symmetry could also violate a Bell inequality that
was properly constructed specifically for this state.

Here is where full quantum relativity can be stated: There is only a single
classically-real correlation contained in the coordinates of a pair of quantum
events.  This correlation can appear in either the negative
($\theta_1-\theta_2)$ or positive ($\theta_1+\theta_2)$ correlation of the
detector coordinates, but not both.  Two quantum events equals one element of
correlation, one element of classical reality.

While the negative singlet wavefunctions  had fermionic character, the positive
singlets have bosonic character---they are symmetric under exchange of
coordinates.  The bosonic or fermionic character of correlated experimental
coordinates depends on the character of the singlet state being measured.

\section{The Dissociated Diatomic}

Consider now an example closer to the original EPR state \cite{EPR}, a
dissociated homonuclear diatomic.  Let $\varphi_n(r)$ be the initial
eigenstate of the bond length coordinate 
\begin{equation}
	r = r_1-r_2
\end{equation}
before dissociation, and $\psi(R)$ be the wavefunction of the center-of-mass coordinate 
\begin{equation}
	R = (r_1+r_2)/2.
\end{equation}
The total wavefunction is then 
\begin{equation}
	\Psi(r,R) = \psi(R)\varphi_n(r).
\end{equation}
Let the diatomic dissociate into two identical neutral atoms, located at $r_1$ and
$r_2$, that subsequently fly apart.  In general, writing out the wavefunction
explicitly in terms of $r_1$ and $r_2$ would yield a non-separable expression
for all but a small set of functions $\phi(R)$ and
$\varphi_n(r)$.  However, the state is trivially always separable in $R$
and $r$ simply by construction, so the wavefunction can be written as
\begin{equation}
	\Psi(r_1,r_2) = \psi(R(r_1,r_2))\varphi_n(r(r_1,r_2)).
\end{equation}
Since it was postulated that the two atoms at $r_1$ and $r_2$ are identical,
which experimentally means that they each can be detected  with the same type
of detector, there should not be a difference if the two detectors are
exchanged.  Hence, the wavefunction must be written as
\begin{eqnarray}\label{EqEPRsinglet}
	\Psi(r_1,r_2) &=& \psi(R(r_1,r_2))\varphi_n( r(r_1,r_2)) \nonumber \\ 
	              & & \pm  \psi(R(r_2,r_1))\varphi_n( r(r_2,r_1)).
\end{eqnarray}
This is now in the form of an excited-state singlet similar to
Eq.(\ref{EqExcitedSinglets}).  Whether the dissociated fragments become a
negative or positive singlet depends on the nature of the initial diatomic.  

As in Figs.\ref{FigNegative} and \ref{FigPositive} for the double-rotor case,
the classical or quantum nature of the atomic fragments following dissociation
will depend on the nature of the singlet state Eq.(\ref{EqEPRsinglet}).  For
all but a small set of $\psi(R)$ and $\varphi_n(r)$, Eq.(\ref{EqEPRsinglet})
will not display Bell correlations and hence $r_1$ and $r_2$ will behave
classically in most cases.  It should be possible experimentally to prepare
various states of the singlet Eq.(\ref{EqEPRsinglet}) by first exciting a beam
of diatomics to specific energy eigenstates to control $\varphi(r)$, passing
the beam through a double-slit to imprint sturcture onto $\psi(R)$, and then
dissociating the diatomics after the slits.

\section{Double-Slit Experiments}

What is generally being measured in double-slit experiments is the correlated 
probability distribution between the initial position $\vec x_i$ at the particle
jet and the final position on a detection screen $\vec x_f$
\begin{equation}
	P(\vec x_f,\vec x_i) = \left\langle \vec x_f \left | \widehat{DS} \right| \vec x_i \right\rangle.
\end{equation}
There are two orthogonal pathways for each possible combination of $\vec x_i$
and $\vec x_f$.  The pathways are "through slit 1, miss slit 2" and "miss slit 1,
through slit 2".  The quantum operator for the double-slit process
can then be written as
\begin{equation}
	\widehat{DS} = | T\rangle |M \rangle \langle T |\langle M | + 
	 	| M\rangle|T \rangle \langle M|\langle T |  
\end{equation}
where $|T\rangle$ means the particle went through the slit, and $|M\rangle$
means the particle missed the slit.  $\widehat{DS}$ can be further factorized
into
\begin{equation}
	\widehat{DS}  = |\Psi_{DS}\rangle\langle \Psi_{DS}|
\end{equation}
where
\begin{equation}
	|\Psi_{DS}\rangle = |T\rangle|M \rangle + | M\rangle| T\rangle  
\end{equation}
is a positive singlet state.  This is the origin of the quantum effects seen in
the double-slit experiment---it is a measurement of a singlet state created by
the slit. Within quantum relativity, all quantum effects comes from an
experimental realization of a singlet state.  

If the screen is placed directly after the slits, then the measured
distribution is
\begin{eqnarray}\label{EqAtSlits}
	P_{near}(\vec x_f,\vec x_i) &=&  \langle \vec x_f | \Psi_{DS} \rangle \langle \Psi_{DS}| \vec x_i \rangle  \nonumber \\ 
			     &=&  \left| \delta(x_s{\rm -}d/2) + \delta(x_s{\rm +}d/2) \right|^2
\end{eqnarray}
where the slits are represented with $\delta$-functions. In practice, finite
slits would imprint a narrow but finite double-peak shape upon the
wavefunction.  When measured with the screen placed far from the slits, the
measured distribution can be written as
\begin{eqnarray}\label{EqFarScreen}
	P_{far}(\vec x_f,\vec x_i) &=&  \langle \vec x_f | \Psi_{DS} \rangle \langle \Psi_{DS}| \vec x_i \rangle  \nonumber \\ 
			     &=& \cos^2\left( \frac{d}{2}\frac{ p_0}{\hbar} \sin \theta \right).
\end{eqnarray}
Both of these measured distributions are characterized only by relative
parameters. In Eq.(\ref{EqAtSlits}), $x_s$ is the position along the axis that
passes through each slit as measured relative to the center of the slits, and
$d$ is the distance between the slits. In Eq.(\ref{EqFarScreen}), $\theta$ is
the angle of the detection position relative to the axis 
that intersects the center of the slits, and $p_0$ is a prior
characterization of the average momentum of the beam that relies on many
relative measurements.  Expressing the measured distribution of quantum
detection events in terms of only relative coordinates of the detection events
is the weak quantum relativistic perspective.  Alternatively, the distribution
Eq.(\ref{EqAtSlits}) can be seen as depending only on the sum ($s{\rm +}d/2$)
and difference ($s{\rm -}d/2$) of detection coordinates---this is the full
quantum relativistic perspective.  One can switch between the weak and full
perspectives with a coordinate transformation.

It is important to note that if the whole double-slit experiment is shifted in
space, no one would expect the probability distributions to change.  The
double-slit experiment does not care about the absolute position of the
experiment.  This is analogous to what is happening in the traditional Bell
scenario.  In Bell experiments there is an  angle invariance that reflects the
fact that the measured probability distribution is independent of absolute
angle.

When trying to measure which slit the particles passed through, the
interference pattern at the far screen is removed.  For example, maybe in the
case of a double-slit experiment with electrons one could flip the spin as it
passes one of the slits but not the other.  Then the detections on the screen
could be correlated with events where the spin-flipper triggers, which would
give information about which slit the electron passed. This experiment is now
described by
\begin{equation}
	P^{(s)}(\vec x_f,s_f,\vec x_i,s_i) = \left\langle \vec x_f,s_f \left | \widehat{DS} \right| \vec x_i,s_i \right\rangle
\end{equation}
where $s_i$ and $s_f$ are the initial and final spin coordinates of the electron,
and the $(s)$ superscript implies this is for the spin-coupled version of the double-slit.
The pathways are now "through slit 1, miss slit 2, no spin flip" and "miss slit 1,
through slit 2, flip spin".  Assuming that the initial spin of the electron beam is 
uniform, the quantum operator for the spin-coupled double-slit process is
\begin{eqnarray}
	\widehat{DS}^{(s)} &=& | T\rangle |M \rangle |\alpha\rangle \langle T |\langle M | \langle\alpha|  \nonumber \\
	 &&+	| M\rangle|T \rangle |\beta\rangle \langle M|\langle T | \langle\alpha|   
\end{eqnarray}
where the $|\alpha\rangle$ and $|\beta\rangle$ are the two possible spin
states, and the incident electron beam was prepared in the $|\alpha\rangle$
state.  
Unlike $\widehat{DS}$, $\widehat{DS}^{(s)}$ is not separable into a process
built from a singlet state.  This results in two possible processes as seen by
the detection screen,
\begin{equation}
	\widehat{DS}^{(\alpha)} = | T\rangle |M \rangle \langle T |\langle M | 
\end{equation}
and
\begin{equation}
	\widehat{DS}^{(\beta)} = | M\rangle|T \rangle  \langle M|\langle T |,
\end{equation}
that do not interfere like in the singlet case but instead represent two
different classical pathways that occur by chance.  From the quantum relativity
perspective, since this process does not require a singlet description it
should not display any quantum effects.

When it is said that quantum particles of a given species are
indistinguishable, what is really indistinguishable are particular dichotomic
pathways that exist in the singlet state being prepared by experiments designed
to measure those quantum particles.  If these pathways are indistinguishable,
then the related relative coordinates being measured behave like quantum
coordinates.  If the pathways are distinguishable through coupling to existing
internal observed variables, then the coordinates behave like classical
coordinates.

\section{Additional Thoughts and Incomplete Speculations}

{\it Relative vs Absolute Reference Frames:}
There seems to have been a discussion between Newton and Leibniz regarding the
nature of space-time.  Newton argued for an absolute coordinate system, a world
stage on which his physical laws played out. Leibniz argued that only
relationships between physical objects exist, that the world is fundamentally
relative.  While special and general relativity as well as quantum field
theory removed the dependence of physical laws from an absolute reference frame
for 4D macroscopic space-time, they still lacked a full relativistic view of
the relationships between microscopic degrees-of-freedom.  They are formulated
in mixed reference frames.  How are the relative and absolute world views
related?  Perhaps what appears local and causally-related and in a relative
world appears non-local and non-causal from an absolute frame and vice versa.

{\it Gravity and QFT:}
Within quantum relativity, gravity could be an emergent phenomena, not
something fundamental.  When we see an object or star with our eyes, we are not
measuring the object or star.  Rather, we are measuring the massless photons
that were coupled to the object or star.  Mass is directly related to the
degrees-of-freedom that we do not see.  This is consistent with $M = E_i(\vec q_i)/c^2$,
where what we call internal energy $E_i$ is expressible in terms of
relative coordinates $\vec q_i$ that are fully internal to the object.  
Massive objects derive their mass from the existence of potentially-observable
but currently-unobserved degrees-of-freedom that they carry.  
Perhaps a photon is effectively massless because it carries no further 
possible internal structure, although this does seem to leave the origin of 
polarization as an open question.

Quantum relativity might help remove infinities in quantum field theory and
general relativity. Maybe no longer a need for renormalization or black hole
core infinities in our descriptions?  Do the infinities in QFT and GR arise
from assuming infinite deegrees-of-freedom somewhere? A quantum relativistic
description can be finite by construction.

Black hole information loss and Hawking radiation could be analogous to burning
a page of writing to destroy the measurable classical information with the
combustion fumes being analogous to the radiation.  The underlying fundamental
degrees-of-freedom of the objects that fall into a black hole can not be
destroyed (quantum information can not be destroyed), but the observed
relations between them (the classically-measurable relationships between the
degrees-of-freedom) are lost.

\acknowledgments{
Thanks to Ben Sussman and Khabat Heshami with whom I've had
many discussions about the subtle aspects of quantum mechanics.}


\begin{thebibliography}{9}

\bibitem{EPR}
A. Einstein, B. Podolsky, and N. Rosen, Phys. Rev. {\bf 47}, 777 (1935).

\bibitem{Bohr}
N. Bohr, Phys. Rev. {\bf 48}, 696 (1935).

\bibitem{Bell}
J. Bell, Physics {\bf 1}, 195 (1964).

\bibitem{CHSH}
J.F. Clauser, M.A. Horne, A. Shimony, and R.A. Holt, Phys. Rev. Lett. {\bf 23}, 880 (1969).

\bibitem{Gabor}
D. Gabor, Nature {\bf 159}, 591 (1947).  

\end{thebibliography}
\end{document}